\documentclass[9pt,preprint2]{aastex}

\usepackage{graphics} 
\usepackage{epsfig} 

\newcommand{\DMU}{$\mathrm{pc} \: \mathrm{cm^{-3}}$} 

\begin{document}

\title{Statistical Studies of Giant Pulse Emission from the Crab Pulsar}

\shorttitle{Giant Pulses from the Crab Pulsar}
\shortauthors{Majid et al.}
\email{walid.a.majid@jpl.nasa.gov}

\author{ 
Walid A. Majid\altaffilmark{1}
Charles J. Naudet\altaffilmark{1}, 
Stephen T. Lowe\altaffilmark{1}, 
Thomas B. H. Kuiper\altaffilmark{1}}

\altaffiltext{1}{Jet Propulsion Laboratory, California Institute of
  Technology, 4800 Oak Grove Dr., Pasadena, CA 91109.}

\begin{abstract}

We have observed the Crab pulsar with the Deep Space Network (DSN)
Goldstone 70~m antenna at 1664 MHz during three observing epochs for a
total of 4 hours.  Our data analysis has detected more than 2500 giant
pulses, with flux densities ranging from 0.1 kJy to 150 kJy and pulse
widths from 125 ns (limited by our bandwidth) to as long as 100 $\mu
s$, with median power amplitudes and widths of 1 kJy and $2\mu s$
respectively.  The most energetic pulses in our sample have energy
fluxes of approximately 100 kJy-$\mu\mathrm{s}$.  We have used this
large sample to investigate a number of giant-pulse emission
properties in the Crab pulsar, including correlations among pulse flux
density, width, energy flux, phase and time of arrival.  We present a
consistent accounting of the probability distributions and threshold
cuts in order to reduce pulse-width biases.  The excellent sensitivity
obtained has allowed us to probe further into the population of giant
pulses.  We find that a significant portion, no less than 50\%, of the
overall pulsed energy flux at our observing frequency is emitted in
the form of giant pulses.

\end{abstract}



\section{Introduction}

While giant pulses (GPs) have been reliably detected from seven
pulsars, (\citealt{knight2006} and references therein) their
properties have only been well studied from two objects, the Crab
pulsar, PSR $\mathrm{B0531+21}$
\citep{sallmen1999,cordes2004,hankins2003,hankins2007}, and PSR
$\mathrm{B1937+21}$ \citep{cognard1996,soglasnov2004}.  In particular,
the Crab pulsar has long been known as a giant pulse emitter. Its
initial discovery \citep{staelin1968} and a number of subsequent
studies have reported remarkable properties of giant pulses from the
Crab.  Giant pulses are a broadband phenomenon (e.g.,
\citealt{sallmen1999}) exhibiting the Galaxy's largest observed
brightness temperature \citep{cordes2004}, and a subset of them are,
in effect, superpositions of extremely narrow nanosecond pulses
\citep{hankins2003,hankins2007}.  It has been shown that energy flux
emission from giant pulses exhibit a power law distribution $N(E >
E_{0}) \sim E^{-\alpha}_{0}$, with $\alpha$ in the range of 1.5-2.5
\citep{popov2007}, in contrast to normal radio pulse emission being
Gaussian distributed \citep{hesse1974}.

In this paper we analyze high time-resolution GP observations from the
Crab pulsar conducted with Goldstone's Deep Space Network (DSN) 70~m
radio telescope at 1.7 GHz (L-band).  With our large data sample, we
carried out a number of statistical studies of GP properties.  We
present radio observations and a description of the recording
strategies and setup in Section 2.  In section 3 we present the data
reduction and analysis scheme. In section 4 we provide a description
of the statistical studies, followed by a discussion of the
implications of the analysis.  Finally, conclusions are given in
section 5.

\section{Observations and Data Set}

As part of an effort to revitalize the L-band system on the DSN 70~m
antenna in Goldstone, a number of pulsar observations were carried out
over four epochs in early to mid 2008.  The Crab pulsar was observed
during two of these epochs for a total observing time of almost four
hours.  For these observations, the front-end electronics
down-converted the L-band RF signal to IF (via an intermediate S-band
up-conversion) for recording, using a pair of VLBI Science Receivers
(VSRs).  The VSRs filter and sample the analog IF signal, then
digitally form sub-channels for recording to disk.  The data used for
this study consist of 32 MHz recorded bandpass in the range 1648-1680
MHz, recorded as four adjacent 8 MHz channels with 2-bit I and 2-bit Q
samples for 8050 continuous seconds, followed 3 min later by two 16
MHz channels with 1-bit I/Q sampling for 1830 continuous seconds.  A
number of hardware problems were encountered during the experiment's
first hour, including antenna pointing and recorder sampling errors.
The data corresponding to these problems were removed.
The configuration of the
channels is summarized in Table 1.


The system temperature during Crab pulsar observations may be
dominated by the emission from the Crab Nebula, one of the brightest
radio sources in the sky.  This is certainly true in our case, where
the nebula is not resolved by the antenna beam.  The nebula is an
extended source with a $\sim5.5'$ diameter and a flux density
parametrized as $S_N \sim 955 \nu^{-0.27} \: \mathrm{Jy}$
\citep{bietenholz1997}, where $\nu$ is the observing frequency in GHz.
At an observing frequency of 1.7 GHz, the nebular flux density is
$S_{N}\sim830 \: \mathrm{Jy}$.  Since the DSN's 70~m antenna beam at
L-band has a half-power width of $\sim8.8'$, the nebula is not
resolved so the system noise must correctly account for the nebular
noise contribution.  The nebular contribution $S_{N}$ is combined with
the contribution from the system temperature in absence of the Crab
Nebula ($S_{sys}^{'}$) to obtain the total system temperature:

\begin{equation}
S_{sys}= S_{N} + S_{sys}^{'}
\end{equation}

On-off measurements of a standard calibrator 3C48, and bright radio
pulsar PSR~B0329+54 were carried out prior to observing the Crab.
These observations yield a system temperature, $S_{sys}^{'}$, of 35~K
(10\% error).  Because the nominal gain $G$ of the 70~m antenna is
$\sim1.0 \: \mathrm{KJy^{-1}}$, the measured system temperature
translates into a system equivalent flux density ($S_{sys} =
T_{sys}/G$) of 35 Jy.  Adding the estimated nebular contribution of
$\sim$~$830 \: \mathrm{Jy}$, yields a total system equivalent flux
density $S_{sys} = 865$~Jy.  This value agrees well with our estimates
obtained from on-off measurements of the Crab, where we measured the
system temperature while alternately pointing at the Crab, and
$1^{\circ}$ away.  We estimate the error for our overall flux density
calibration scale to be less than $\sim20\%$.  Our detection
sensitivity for single pulses is determined using the radiometer
equation $\Delta S = \eta S_{sys}/\sqrt{\Delta \nu \Delta t}$, where
$\eta$ is the digitization loss factor, $\Delta \nu = 32 \:
\mathrm{MHz}$ is the recorded bandwidth and $\Delta t = 0.1 \:
\mathrm{\mu s}$ is the minimum sample time.  With two-bit digitization
we have $\eta = 1.3$ and we obtain a $1 \sigma$ detection threshold of
$\Delta S_{min} \sim$~560~Jy.  For a $7\sigma$ detection threshold
the minimum single-bin detectable pulse
amplitude is expected to be 3.9~kJy.  For comparison, Table 2 lists
the parameters of previous Crab GP studies with the current work.

\begin{deluxetable}{ccccccccc}
\footnotesize
\tablecaption{VSR channel configuration \label{tbl-1}}
\tablewidth{0pt}
\tablehead{
\colhead{Epoch Number} &\colhead{NRec\tablenotemark{a}} &
\colhead{Chan 1}   & \colhead{Chan 2} &
\colhead{Chan 3}   & \colhead{Chan 4} &
\colhead{BW (MHz)\tablenotemark{b}} &
\colhead{Nbits\tablenotemark{c}} &
\colhead{Obs Time (min)}
}
\startdata
1 & 2 & 1652 & 1660 & 1668 & 1676 & 8  & 2 & 135 \\
2 & 1 & 1656 & 1672 &   -  &  -   & 16 & 1 & 30  \\
3 & 2 & 1652 & 1660 & 1668 & 1676 & 8  & 2 & 75  \\

\enddata

\tablenotetext{a}{Number of recorders}
\tablenotetext{b}{Channel Bandwidth}
\tablenotetext{c}{Number of recorded bits per sample}

\end{deluxetable}

\begin{deluxetable}{ccccccccc}
\tabletypesize{\footnotesize}
\tablecaption{Crab pulsar giant pulse observations \label{tbl-2}}
\tablewidth{0pt}
\tablehead{
\colhead{$\nu (GHz)$\tablenotemark{a}} &
\colhead{Epoch (MJD)} &
\colhead{T (hr)} &
\colhead{BW (MHz)} &
\colhead{$\Delta T$ ($\mu s$)}  &
\colhead{SEFD\tablenotemark{b}} &
\colhead{Ndet\tablenotemark{c}} &
\colhead{Telescope} &
\colhead{Reference}
}
\startdata


0.1-0.2   & 53635       & -    & 6    & 1024-256 & 1100 &  31    & MWA-LFD   & \citep{bhat2007} \\
0.430     & 52304       & 1.0  & 12.5 & 128      & 1262 &  11880 & Arecibo   & \citep{cordes2004} \\
0.6/1.4   & 50224       & -    & 50   & 1        & -    &  29    & VLA       & \citep{sallmen1999} \\
0.812     & 48433       & 100  & 20   & 200-300  & 13.5 &  $3 \times 10^{4}$ & GB 43m & \citep{lundgren1995} \\
1.18      & 52277       & 0.47 & 100  & 100      & 309  &  863   & Arecibo   & \citep{cordes2004} \\
1.197     & 52944       & 3.5  & 20   &  4       & 215  &  17869 & WSRT      & \citep{popov2007} \\
1.475     & 52277       & 0.58 & 100  & 100      & 291  &  647   & Arecibo   & \citep{cordes2004} \\
1.3-1.47  & 53736       & 3    & 64   & 0.5      & 1100 &  706   & ATCA      & \citep{bhat2008} \\
1.7       & 54618       & 3    & 32   & 0.1      & 865  & 2500   & Goldstone & This paper \\
2.15      & 52304-52306 & 0.15 & 100  &  32      & 79   &  135   & Arecibo   & \citep{cordes2004} \\
2.33      & 52315       & 0.15 & 100  &  32      & 78   &  92    & Arecibo   & \citep{cordes2004} \\
2.85      & 52306       & 0.26 & 100  &  32      & 74   &  103   & Arecibo   & \citep{cordes2004} \\
3.5       & 52398-52412 & 1.27 & 100  &  64      & 41   &  549   & Arecibo   & \citep{cordes2004} \\
4.15      & 52295-52337 & 1.49 & 100  &  32      & 20   &  1663  & Arecibo   & \citep{cordes2004} \\
4.5-10.5  & 53005-53736 & -    & 2200 & 0.0004   & -    &  380   & Arecibo   & \citep{hankins2007} \\
5.5       & 52336-52411 & 0.3  & 100  &  32      &  20  &  22    & Arecibo   & \citep{cordes2004} \\
8.8       & 52398-52414 & 1.42 & 100  &  16      & 22   &  2249  & Arecibo   & \citep{cordes2004} \\

\enddata

\tablenotetext{a}{For comparison convenience this table is sorted by
  lowest observing frequency in ascending order}
\tablenotetext{b}{System equivalent flux density (Jy)}
\tablenotetext{c}{Number of detected GPs}

\end{deluxetable}

\section{Data Reduction}

Our 4 hours of Crab pulsar data, consisting of ~540 Gigabytes, were
recorded to disk and shipped to JPL for post-processing.  The average
and RMS RF voltage was computed for each second of data as a quick
assessment of data quality.

\subsection{Coherent Dedispersion and Normalization}

The data were coherently dedispersed using the nominal dispersion
measure (DM) value\footnote{The DM value was obtained
  from Jodrell Bank Crab pulsar monthly ephemeris:\\
  http://www.jb.man.ac.uk/$\sim$pulsar/crab.html.}
 of 56.7671 \DMU, following the
dispersion removal technique developed by \citet{hankins1975}.  Each
IF channel was dedispersed separately by performing an FFT on a full
second of complex (I and Q) samples, plus the fractional second of
data required to fill the FFT arrays with a power of two number of
samples.  For the 2-bit, 8 MHz data, the approximately 1 Hz frequency
bins were counter-rotated in phase to remove the frequency-dependent
dispersive delay, then inverse-transformed back to the time domain.
The 1-bit, 16 MHz data were processed similarly except the final
inverse FFT was performed on each half of the channel bandpass
separately, effectively splitting each 16 MHz channel into two 8 MHz
channels.  In this way, all data could be further processed in a
similar manner.  Dedispersing a full second of data kept delay
smearing over each $\sim1$~Hz frequency bin much less than a sample,
ensuring the dedispersed data maintained its full time resolution.

After dedispersion, we form each sample's normalized power, $P_i$,
averaged over all four frequency channels, where $i$ is the sample
number.  This was done by independently normalizing each complex
component (I and Q) from all four frequency channels so that all eight
quantities had zero mean and unit standard deviation.  These eight
quantities were squared and averaged to form our power time series,
using:
\begin{equation}\label{sum_power}
P_i = \frac{1}{8}\sum_{k=1}^4 (I_k^{2} + Q_k^{2}),
\end{equation}
where $k$ runs over the four frequency channels.  Assuming each I and
Q component is normalized and Gaussian distributed, in the no-signal
limit $P_i$ will have a {\it reduced} chi-squared distribution with
eight degrees of freedom.

\subsection{Pulse Detection}

Our pulse-detection algorithm begins with the 8~MHz $P_i$ time series
whose noise component is nominally modeled by the reduced chi-squared
distribution with 8 degrees of freedom.  In general, with $\nu$
degrees of freedom, the reduced $\chi^2$ probability density function,
parameterized by $x$, is given by:

\begin{equation}\label{prob_chi2}
{\mbox{pdf}(x) = \frac{\nu {(\nu x)}^{(\nu-2)/2}e{}^{-\frac{\nu x}{2}}}{2^{\nu/2}\Gamma{(\nu/2)}}},
\end{equation}

where the $\Gamma$ function for integer and half-integer arguments is 
defined as:

\begin{equation}\label{gamma}
\Gamma(n) = (n - 1)!,~~ \Gamma(\frac{1}{2}n) = \frac{(n - 2)!!\sqrt{\pi}}{2^{(n - 1)/2}}.
\end{equation}

\begin{figure*}[h!t]
\centerline{\epsfig{file=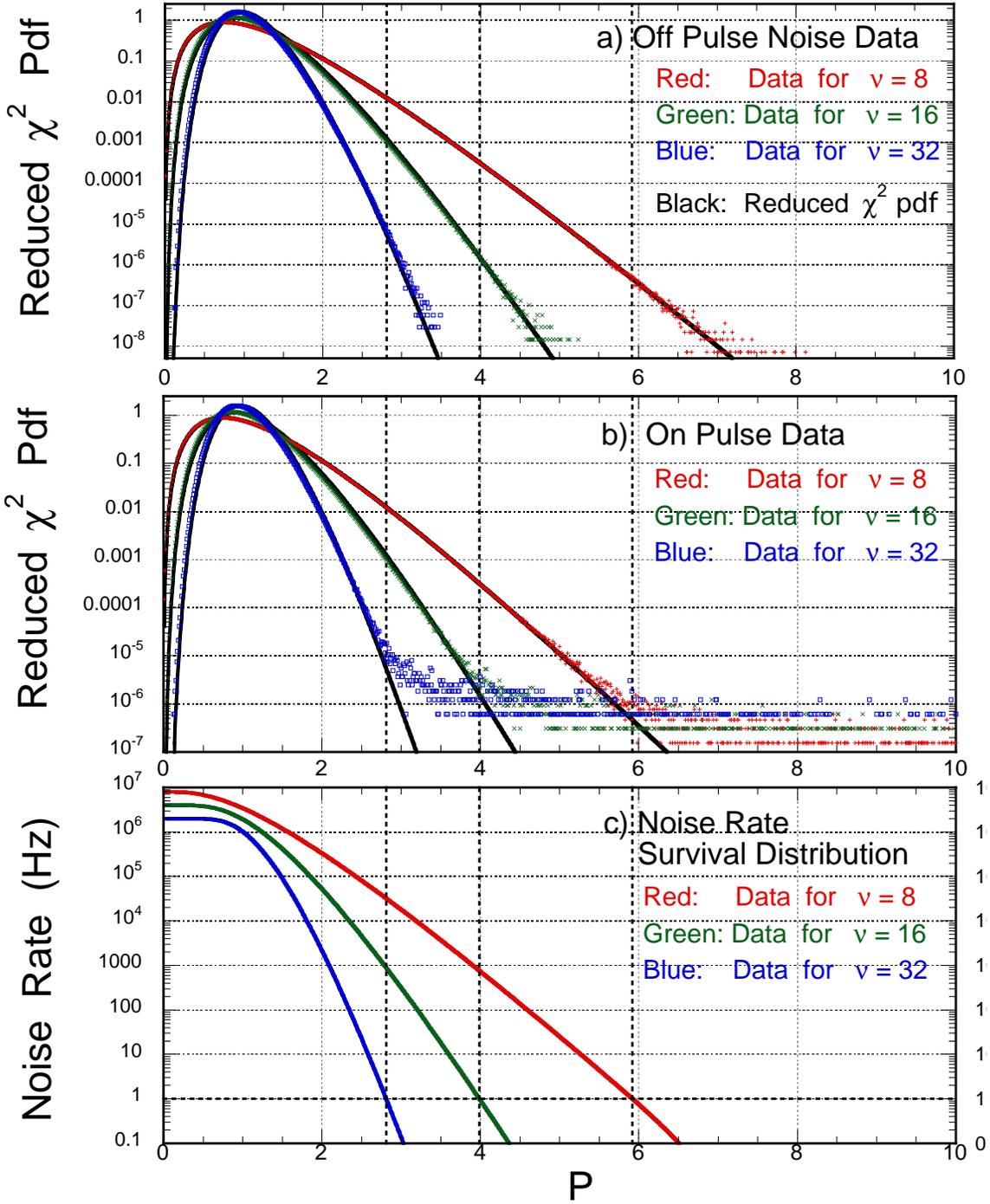, height=7.25in}}
\caption{a) The background pulse $\chi^{2}_{\nu}$ distribution for
  $\nu$ of 8, 16 and 32; the red, green and blue data points
  respectively. b) The background pulses plus giant pulse
  $\chi^{2}_{\nu}$. The corresponding theoretical $\chi^{2}_{\nu}$
  distributions are the solid black curves. c) The noise rate survival
  distribution for $\nu$ of 8, 16 and 32; the red, green and blue
  curves respectively. The black horizontal dashed line, at a noise
  rate of 1 Hz, intersects the three $\chi^{2}_{\nu}$ distributions at
  the corresponding power cut points. Three black vertical dashed power
  cut lines have been extended through all three figures.  }
\label{fig:chi_squared}
\end{figure*}

The red data points in Figure \ref{fig:chi_squared}a show the measured
$P_i$ distribution for 30 minutes of data.  The data samples were
taken during the Crab pulsar off-pulse regions of the pulse phase.
The black solid curve in \ref{fig:chi_squared}a is the corresponding
theoretical expectation given by Equation~\ref{prob_chi2} with
$\nu=8$.  Excellent agreement is seen in over eight decades.

In the time domain, our matched filter algorithm averages $N$
temporally consecutive $P_i$ power samples, and compares this with a
threshold parameter designed to strongly reject noise while passing
GPs with high efficiency. The filter used in this analysis begins with
the $P_i$ power series corresponding to $N$=1. The $N$=2 time series
is formed by averaging each consecutive pair (without overlap) of the
$N$=1 series, resulting in half the number of power samples.  This
doubling procedure is repeated up to the creation of the $N$=2048
power series which corresponds to 256~$\mu$s wide time-averaged bins.
For each power series, summing over the $N$ samples results in:
\begin{equation}
{P_i^N = \frac{1}{N}\sum_{m=0}^{N-1} P_{i+m}}
\end{equation}
where $i$ is in the range $(0, N, 2N, ...)$.  Because each $P_i$
averages 8 Gaussian squares, $P_i^N$ will have its noise component
distributed according to Equation~\ref{prob_chi2} with $\nu=8N$,
assuming the data samples $P_i$ are truly independent.\footnote{As $N$
  increases, a small sample-to-sample correlation was seen.  This loss
  of true data independence was handled by making a small correction
  to the ideal $\nu$. For example, the $N_s=4$ case, ideally $\nu=32$
  but $\nu$=31 yields a better fit to the data.}
Figure~\ref{fig:chi_squared}a shows the off-pulse power, $P_i^N$, for
$N=1$ (red points $\nu=8$ in Equation~\ref{prob_chi2}), $N=2$ (green
curve, $\nu=16$) and $N=4$ (blue curve, $\nu=32$). The black curves
give the theoretical distributions from Equation~\ref{prob_chi2} in
each case.

In order to prevent cut-induced biases in the giant-pulse width
distribution, proper threshold cuts should result in the same number
of background noise events per second, on average, regardless of $N$.
We define $R$ to be the rate of noise fluctuations expected to pass a
threshold cut $\xi(N)$, in units of events per second.  This is
equivalent to picking a constant event confidence level independent of
pulse width. The noise rate $R$, also the false alarm rate, is equal
to the probability that an averaged power computed with $N$ samples
exceeds $\xi$, times the number of trials in 1 second:
\begin{equation}\label{eq_Rate}
R(\xi) = S(\xi)\times \frac{f_s}{N}
\end{equation}
where $S(\xi)$ is 1 minus the cumulative probability distribution
corresponding to Equation~\ref{prob_chi2}, also known as a survival
function, and $f_s$ is the number of samples per second.
Figure~\ref{fig:chi_squared}c shows the the noise rate as a function
of $P$ for the $N=1$ (red points), $N=2$ (green curve, $\nu=16$) and
$N=4$ (blue curve, $\nu=32$) cases.

In the limit of large $\xi$, reasonable given our desire to strongly cut noise, 
this becomes:
\begin{equation}\label{eq_R}
R =
\frac{f_s}{N}
\frac{(4N\xi)^{4N-1}e^{-4N\xi}}{(4N-1)!}.
\end{equation}
The threshold values are computed from
this equation iteratively using
\begin{equation}\label{eq_eta}
\xi(N) = \frac{1}{4N} \left [(4N-1)\log(\xi)-\log(R)+C \right ]
\end{equation}
where
\begin{equation}
C=(4N-1)\log(4N) + \log(\frac{f_s}{N}) - \log((4N-1)!)
\end{equation}
For each averaged power $P_i^N$ in every power series, Equations
\ref{eq_R} and \ref{eq_eta} are used to compute the effective noise
rate $R$ corresponding to the measured $P_i^N$.  This effective noise
rate estimates the likelihood that a given averaged power value is a
noise fluctuation.  The dotted horizontal line in
Figure~\ref{fig:chi_squared}c is the 1 Hz rate cut, $R$-cut=1.0 Hz,
and the corresponding power cuts $\xi(N=1)$, $\xi(N=2)$ and $\xi(N=4)$
are shown as the three vertical dotted lines in
Figure~\ref{fig:chi_squared}a and Figure~\ref{fig:chi_squared}b.  As
described below, we reduce background noise by cutting on $R$, using
the $\xi(N)$, in order to guarantee that the background rate is
independent of $N$.

This consistent accounting of the probability distributions and
equivalent threshold cuts to keep background rates independent of
pulse width ensures balanced detection efficiencies even for extremely
large-width giant pulses.

Figure~\ref{fig:chi_squared}b plots the $\chi^{2}_{\nu}$ distributions
for all the data, so as to include the Crab's giant pulses: the long
tail of events with large $\chi^{2}_{\nu}$ is clearly seen.  With each
detection, both $N$ and the probability of such a pulse being a noise
fluctuation, expressed as $R$ in Equation~\ref{eq_R}, are stored.

This pulse-detection algorithm typically detects each pulse candidate
a large number of times, so a method is needed to transform these
multiple detections into individual pulse candidates, and to estimate
their properties such as time, total power, and width.  The algorithm
used in this analysis was to flag any sample that participates in a
detection, regardless of $N$, then call any contiguous interval of
flagged samples a pulse candidate.  A number of parameters were
computed for each pulse candidate.  The peak power, the pulse
integrated power, the time at peak power, the power-weighted mean
pulse time, and the power variance about the mean time, were all
computed from the 8 MHz channel-summed powers.  From the multiple
detections associated with each pulse candidate the lowest-probability
$R$, corresponding to the highest significance, was chosen as the
giant pulse.  The associated $N$, power, and time were then saved.

\begin{figure*}[bp]
\centerline{\epsfig{file=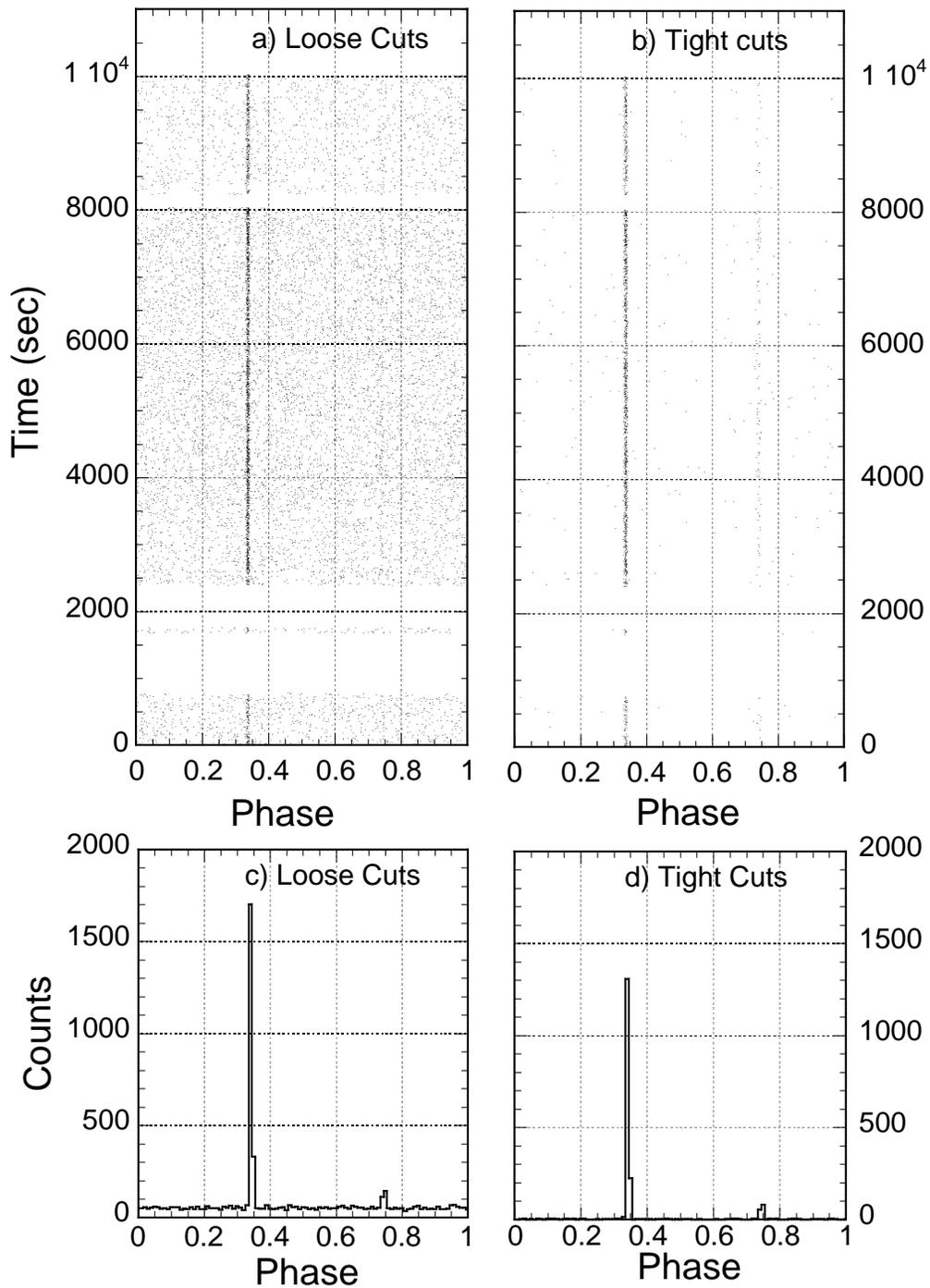, height=7.25in}}
\caption{Time phase scatterplots of the detected pulses for the loose
and tight data sets. The main and interpulse giant pulses are 
seen as straight lines at phases of 0.337 and 0.737 respectively. 
The bottom histograms shows the corresponding
phase projection. The improvement in signal to noise is clearly
seen with the tight cuts.}
\label{fig:time_phase}
\end{figure*}

Figures \ref{fig:time_phase}a and \ref{fig:time_phase}b show
scatterplots of pulse phase and time for all giant pulses found in our
data set, for both a relatively soft ($R$=0.02 Hz) and hard cut
($R$=0.0006 Hz), respectively\footnote{The significance of these cut
  values becomes clear later in the paper.}.  The main pulse, with
phase near 0.3, and the interpulse, near 0.7, are clearly evident.
The loose R-cut (0.02 Hz) implies more white-noise contamination due
to the corresponding lower ${\chi}^{2}_{\nu}$ thresholds.  Similarly,
the tight R-cut (0.0006 Hz) reduces noise contamination on higher
${\chi}^{2}_{\nu}$, which is equivalent to a higher power
signal-to-noise (SNR) threshold.  The projections of these
scatterplots onto the time axis, the pulse phase histograms, are
presented in Figures \ref{fig:time_phase}c and \ref{fig:time_phase}d
for the hard and soft cut, respectively.  In the loose-cut case, over
1600 main giant pulses (GPs) are seen on a flat background of ~100
noise pulses.  With tight cuts, over 1200 main pulse GPs are found
with no background, and a clear peak of GPs is seen at the interpulse
phase region.  By defining an on-pulse phase range of (0.32, 0.35) for
the main pulse phase, and (0.72, 0.75) for the interpulse phase, we
can study the count rate and SNR of the GPs as a function of $R$-cut.


If $S_T$ is defined as the number of counts within the on-pulse region
and $S_N$ is the number of background counts estimated from the
off-pulse phase region, the background-subtracted signal count is
$S=S_T-S_N$.  Assuming counting statistics, the SNR is then:

\begin{equation}
SNR =(S_T-S_N)/(S_T+S_N)^{1/2} \end{equation}

\begin{figure}[h!t]
\centerline{\epsfig{file=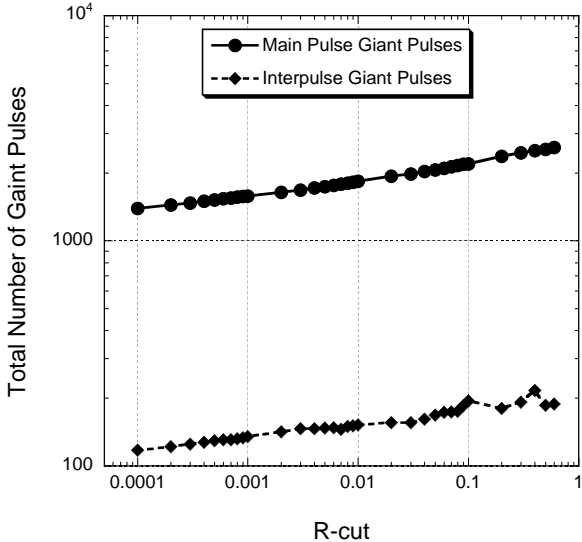, width=\linewidth}}
\caption{The number of background-subtracted Giant Pulses, for both
the Main pulse and Interpulse, as a function of the noise cut,
$R$-cut. As the $R$-cut increases (power amplitude cut decreases) the
total number of giant pulses increase.}
\label{fig:numgp_rcut}
\end{figure}

Figure \ref{fig:numgp_rcut} shows the total number of giant pulses as
a function of $R$-cut, for both the main and interpulse regions.  For
the loosest cut processed ($R$=0.6), over 2500 main phase GPs and 200
inter phase GPs are observed.

Careful examination of the pulse-candidates in Figure
\ref{fig:time_phase}a shows density variations during the experiment's
first and last hours.  Our detection thresholds were adjusted to allow
an approximately constant noise rate as a function of pulse width,
however Figure \ref{fig:time_phase}a shows that the noise rate is not
constant in time.  At the experiment's start, there were a number of
problems with the front-end sampler voltage thresholds, which reduced
the data sensitivity.  The single-bit data beginning at time 8230~s
also shows a low pulse-candidate density.  Even though dispersion
mitigates the 1-bit sampling loss to some extent, as explained above,
these data are somewhat less sensitive than the 2-bit data.  Efforts
to reconcile the different sampling effects and resulting
sensitivities, for example by scaling the sample values or modifying
the detection thresholds, significantly increases the complexity of
the analysis.  Thus, from this point forward the analysis is
restricted to only those data taken from time 2400 to 8050 s, a total
of 5650 seconds of data, as these data have the highest sensitivity,
and show an approximately constant sensitivity.

\begin{figure}[h!t]
\centerline{\epsfig{file=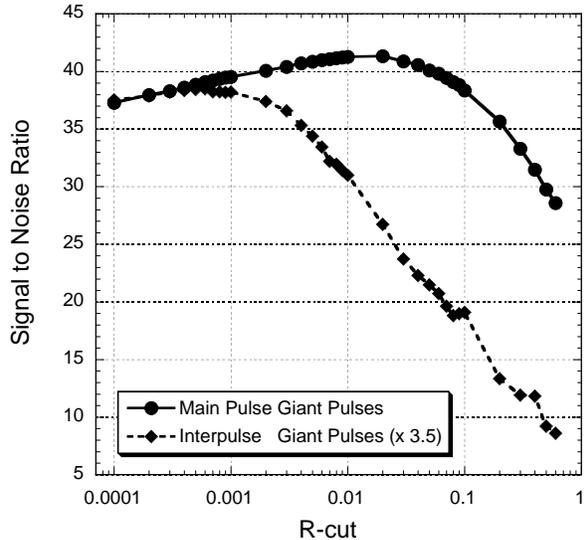, width=\linewidth}}
\caption{The giant pulse signal-to-noise ratio as a function of the
noise cut, $R$-cut, for both the main pulse and interpulse GPs.}
\label{fig:snr_rcut}
\end{figure}

Figure \ref{fig:snr_rcut} shows the SNR as a
function of $R$-cut for both the main and interpulse pulses.  It is seen
that the maximum SNR for the main pulses are at $R$ = 0.02 and
$R$=0.0006 respectively.
lowering the $R$-cut (increasing the total $\chi_{\nu}^2$ cut) fewer
GPs with higher SNR values are detected.  In this paper we will refer
to three cuts, a loose cut at $R$=0.02, a tight cut at $R$=0.0006, and
the loosest cut at $R$=0.6 resulting in the largest sample of Main
Pulse (MP) GPs found in this analysis ($\simeq$2500).  Yet even with the
tight cut over 1200 MP GPs and over 100 Inter-Pulse (IP) GP are
detected both without any significant background contamination.

\section{Various Statistics of Detected Giant Pulses}

The tremendous amount of energy radiated by giant pulses is observed
with a wide variety of pulse morphologies.  One of the goals of
giant-pulse research is to determine, as closely as possible, the
originating pulse shape, its amplitude and width.  A wide variety of
factors influence the shape of the pulse, such as dispersion,
instrumental smearing, interstellar scintillation and scattering due
to turbulent media.  These effects all tend to broaden a pulse,
decreasing its true peak amplitude and increasing its width.  In our
data set, the amplitude is simply defined as the peak power, or rather
the peak flux density, found in the optimum smoothed data set.  Past
attempts to characterize pulse widths for a large sample of GPs have
been limited either to coarse sampling size or large dispersive
smearing and small sample sizes
\citep{lundgren1995,popov2007,bhat2008}.  Our L-band observations,
employing coherent dedispersion on large bandwidth baseband channels
minimize the dispersive smearing, obtaining a 125 $\mu$s time
resolution, contain over 1200 GPs, which should allow more robust
width measurements.

\subsection{Pulse Amplitudes and Widths}

\begin{figure*}[bp]
\centerline{\epsfig{file=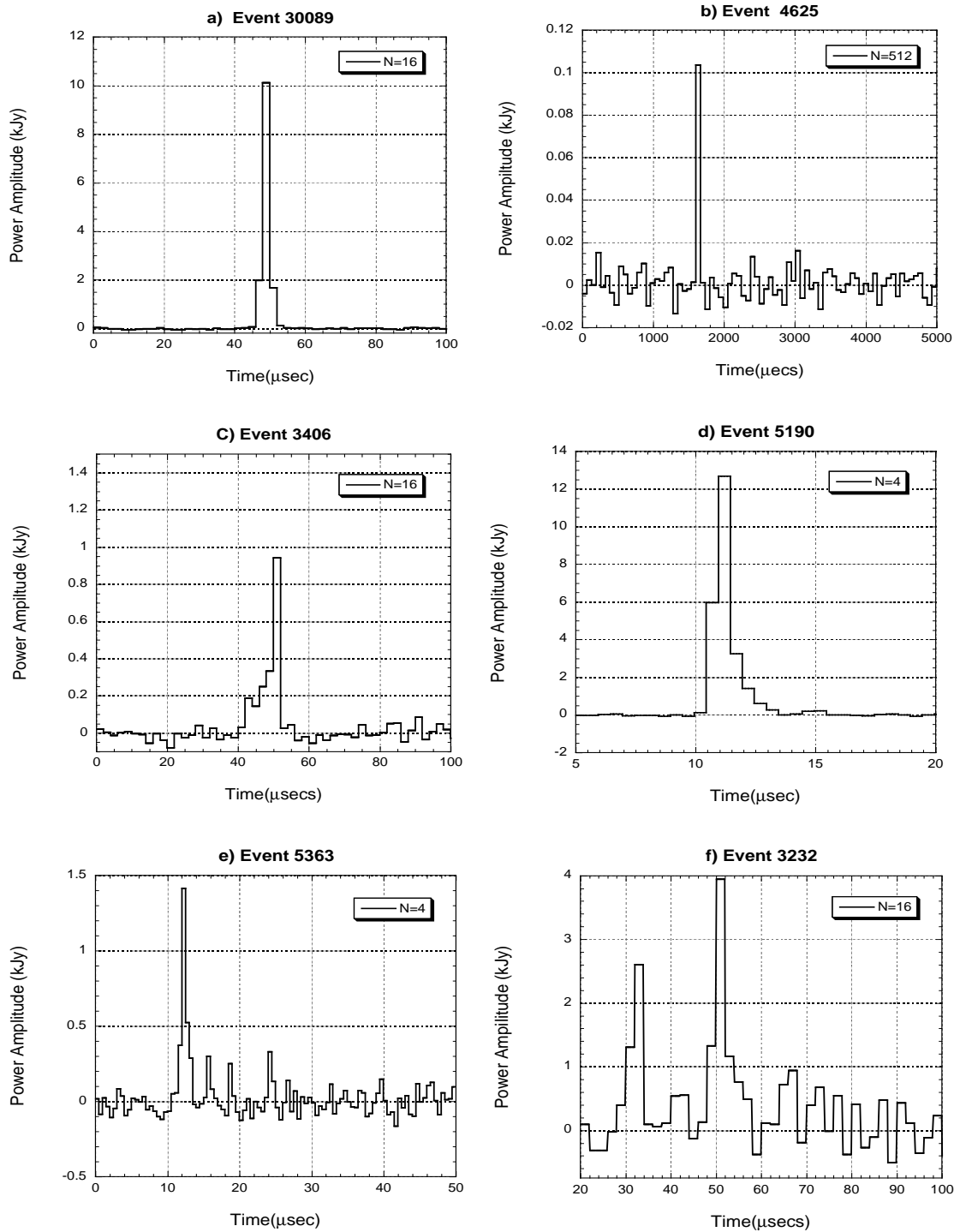, height=7.5in, width=5.8in}}
\caption{Giant pulse time series plots. Each is shown with it's
optimum smoothing width "N". The widest pulse shown is pulse b with
N=512 and the two narrowest shown are pulses d and e with N=4}
\label{fig:pulses}
\end{figure*}

Visual examination of many pulses, some of which are shown in Figures
\ref{fig:pulses}a-f, reveals that variations in pulse morphology
represent the dominant systematic error in pulse amplitude and width
determinations.  The six GP events shown are plotted using their
optimum smoothing, with widths ranging from the narrowest (smoothing
widths of $N$=4) to the widest ($N$=512), and with amplitudes ranging
from 1 kJy to above 100 kJy.  Although a large subset of GPs can be
well fit to a Gaussian shape, an equally large sample of GPs show
non-Gaussian shapes, as presented in Figures \ref{fig:pulses}c-f.
Given a significant fraction of GPs having non-Gaussian shapes,
fitting GPs with a Gaussian model yields poor chi-squared fits, and
thus poor amplitude and width estimates.

\begin{figure*}[bp]
\centerline{\epsfig{file=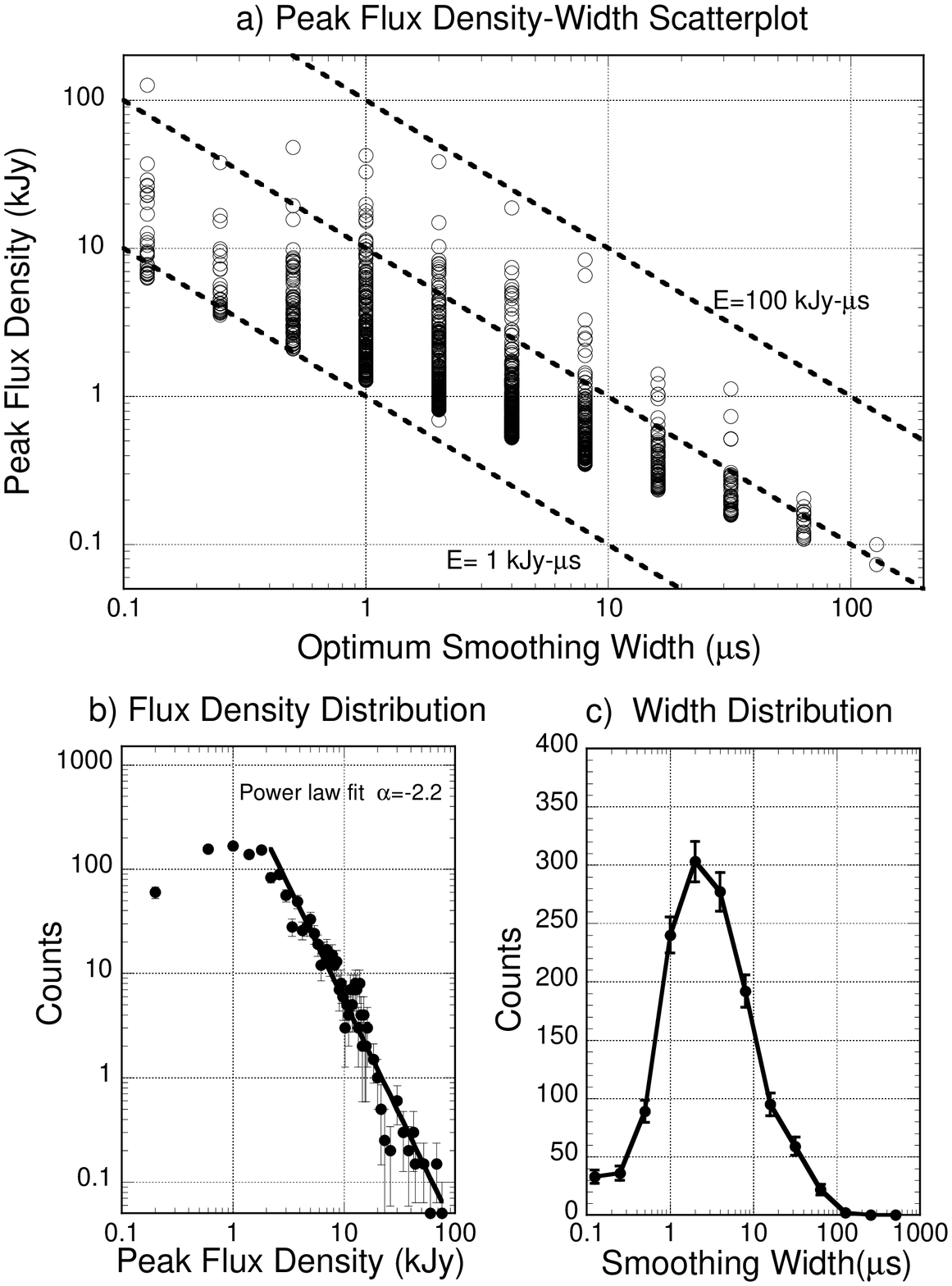, width=5.5in}}
\caption{The scatterplot of peak flux density and optimum smoothing width
($W_{s}$) along with the corresponding peak flux density (b) and width projection
histograms (c) for the tight cut. The diagonal dotted lines in the scatterplot (a)
are energy isotherms
at E= 1 kJy-$\mu$s, 10 kJy-$\mu$s and 100 kJy-$\mu$s}
\label{fig:amp_width}
\end{figure*}

An alternative width definition is simply the optimum smoothing width
$W_{s}=N\Delta t$ ($\Delta t$ is the intrinsic sample resolution)
obtained from the giant pulse detection algorithm.  The amplitude and
timing can then be defined using the bin with the greatest power
amplitude.  With these alternative definitions of power amplitude and
width, pulse morphology can be seen in Figure \ref{fig:amp_width}a,
where the scatter plot shows GP peak flux density versus width, for
all the GPs passing the tight-cut.  A strong correlation between the
observed peak flux density and smoothing width is observed, with
higher peak flux density seen at lower smoothing widths.  The diagonal
dotted lines in Figure \ref{fig:amp_width}a show iso-energy flux
contours of 1 kJy-$\mu$s, 10 kJy-$\mu$s and 100 kJy-$\mu$s.  A wide
variation of shapes can be seen for the same observed energy flux.
For example, a large pulse with E = ~10 kJy-$\mu$s can be very sharp,
less than a $\mu$s in width with over 10 kJy peak flux density, or
very broad, with widths over 10 $\mu$s but flux density less than 1
kJy.  Although great care was taken to ensure that no pulse width
selection bias occurred, a comparison of the GP data with the iso-energy
flux contours suggests that giant pulses with total energy fluxes of $\sim$1
kJy-$\mu$s cannot be extracted above smoothing widths of 2 $\mu$s.

Figures \ref{fig:amp_width}b and \ref{fig:amp_width}c show the
corresponding amplitude and width projections.  The peak flux density
distribution in Figure~\ref{fig:amp_width}b indicates the median pulse
flux density is $\sim$1 kJy with a minimum observed peak flux density
of $\sim$0.1 kJy and a maximum of $\sim$100 kJy.  Above a peak flux
density of 2 kJy, the histogram shows a power-law dependence out to
~100 kJy.  A power-law fit yields an exponent of $\alpha$ = -2.2 +/-
0.1, very close to that found by \citep{bhat2008}, where at slightly
lower frequencies of 1300 MHz and 1460 MHz, values of -2.3 and -2.2
were measured respectively.

The distribution of the smoothing width, $W_{s}$, is shown in Figure
\ref{fig:amp_width}c.  The median $W_{s}$ is 2 $\mu$s and the mean
$W_{s}$ is $\sim$6 $\mu$s with a minimum of 0.2 $\mu$s with a long
tail extending out to 100 $\mu$s.  This width range compared to
previous Crab GP datasets \citep{bhat2008} is extended to pulses with
10 times wider and two times narrower widths.

\begin{figure}[h!t]
\centerline{\epsfig{file=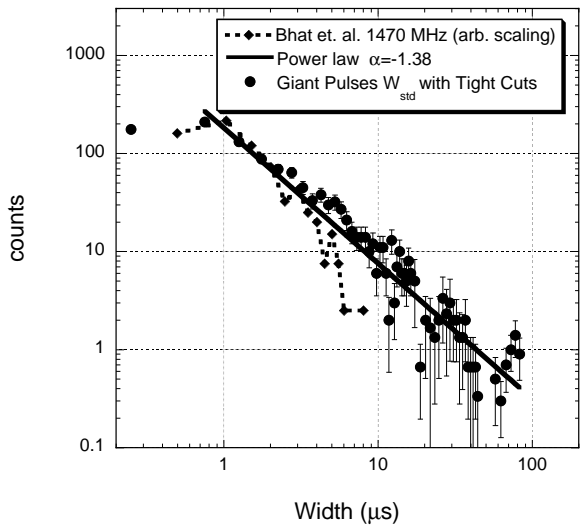, width=\linewidth}}
\caption{The giant pulse width histogram for the tight data set.
The fitted power law is indicated by the solid line. 
The dotted histogram is data of \citep{bhat2007}}
\label{fig:width}
\end{figure}

Although the smoothing width, $W_{s}$ is a convenient definition of
pulse width, it does not always appear to be optimum.  Sometimes, as
shown in Figure \ref{fig:pulses}b, a wide giant pulse with $N$=512 has
almost all its energy within the one smoothed bin.  Yet as Figures
\ref{fig:pulses}c-f illustrate, a large fraction the time signifiant
energy is spread over many bins at the optimum smoothing width.
Following others \citep{bhat2008} we define an ``effective pulse
width,'' $W_{std}$, which is the standard deviation (std) for all
``acceptable'' bins in a giant pulse, where acceptable bins are
defined with a simple algorithm.  Raw data with the highest time
resolution is used, and data within $\sim$3$N$ of the maximum power
amplitude sample is selected.  Only bins with power amplitudes greater
than 2 sigma above the background power are included in the first
estimation of the $W_{std}$, and the background level.  The final
calculation sums over all bins within $\sim$3$W_{std}$ of the peak,
again with bins having power amplitudes greater than 1 sigma above the
background.  This algorithm allows us to estimate the pulse standard
deviation and higher order moments such as skewness and kurtosis.
This procedure gives an estimate of the pulse width ($W_{std}$) that,
as expected, gives excellent agreement with the fitted Gaussian sigma
and gives a reasonable estimation of the pulse width in the
non-Gaussian cases.  A comparison with pulse widths extracted by Bhat
et. al., at 1470 MHz, can be seen in Figure \ref{fig:width}.  Due to
the higher time resolution in this study, pulses with widths a factor
of 2 narrower are detected.  Perhaps the more significant difference
with the earlier data of Bhat et. al. is the increase in the
sensitivity at very large pulse widths.  Bhat et. al. observed no
pulses with widths greater than 10 $\mu$s, whereas the present data
set has a significant number of pulses extending out to over 100
$\mu$s with a median $W_{std}$ of 2.35 $\mu$s and a mean of 10.45
$\mu$s.
We have parameterized this distribution with a good power law fit, 
$N(W_{std}) \propto W_{sd}^{-1.3}$, that extends from $\sim$ 1 $\mu$s to almost
100 $\mu$s. 

\subsection{Pulse Energies}

\begin{figure}[h!tb]
\centerline{\epsfig{file=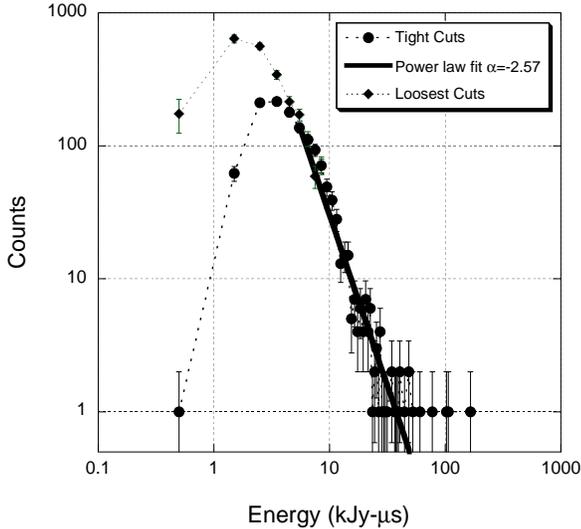, width=\linewidth}}
\caption{The giant pulse energy flux distribution with a power law
fit of $\alpha$ = -2.57 for energy fluxes between 5 and 50 kJy-$\mu$s.}
\label{fig:energy}
\end{figure}

The measurement of giant pulse parameters such as power amplitude,
width, and energy flux are critical in constraining models of GP
emission.  The total pulse energy flux has the advantage, as compared
to pulse width, of being a robust pulse estimation parameter, as it is
relatively insensitive to pulse shape.  Also, since the total pulse
energy flux is a simple power sum over the pulse duration, the random
background noise will tend to average down to zero.  The GP energy
flux distribution, for the tight cut data set, is shown as black
circles in Figure \ref{fig:energy}. Energy fluxes up to $\sim$100
kJy-$\mu$s are found, consistent with earlier studies
\citep{bhat2008}. The average observed GP energy flux for all the main
giant pulses was $\sim$7 kJy-$\mu$s with a median energy flux of
$\sim$5 kJy-$\mu$s.  Above 5 kJy-$\mu$s the data are well fit by a
power law, the solid line in Figure \ref{fig:energy} shows fit to the
data with $\alpha$=-2.57.  A very sharp energy flux threshold is
observed at $\sim$3 kJy-$\mu$s. The black diamonds in Figure
\ref{fig:energy} show the main GP energy flux distribution for the
loosest cut examined. An additional $\sim$1400 main GPs are found but
all at lower energy flux as would be expected. The additional main GPs
seem to follow the same power law as the data with tight cuts but the
energy threshold is reduced to $\sim$1.5 kJy-$\mu$s. No evidence is
seen for a roll over or softening of the intrinsic power law.  The
turnover in Figure \ref{fig:energy} is most likely due to the applied
threshold cut.

Following a previous analysis \citep{knight2006}, the cumulative
energy flux distribution, defined as the probability per second of a
pulse having an energy flux $E$ greater than a value $E_0$, is used to
compare the occurrence frequency of giant pulses between different
experiments:

\begin{equation}
P(E > E_0) = kE^{\alpha}\end{equation}

\begin{figure}[h!t]
\centerline{\epsfig{file=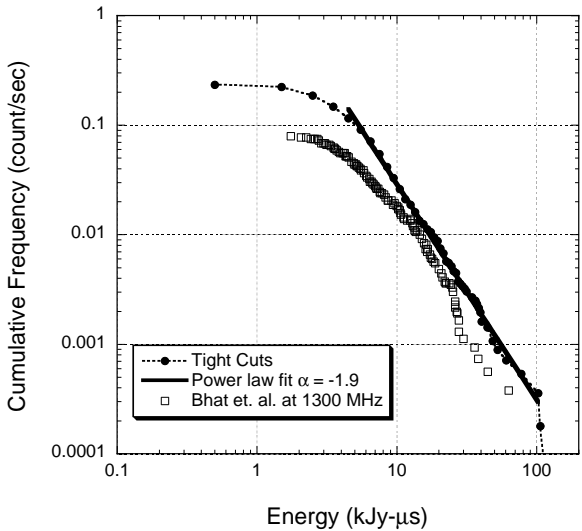, width=\linewidth}}
\caption{The cumulative energy flux distribution for giant pulses. A power
law fit for energies above 5 kJy-$\mu$s yields $\alpha$=-1.9.}
\label{fig:summed_energy}
\end{figure}

Figure \ref{fig:summed_energy} shows the observed cumulative energy
flux distribution for the tight data set, solid black circles, along
with the data from Bhat et. al., open squares.  The solid line is a
power-law fit to the data in the energy range 5 to 100 kJy-$\mu$s.
Above the energy flux threshold, the data are well fit by a power law
and no evidence is found for high-energy flux cut-offs.  The best-fit
power law exponent is found to be $\alpha$=-1.9 +/- 0.05 The agreement
with Bhat et. al. is very good in the energy flux region of 10-20
kJy-$\mu$s but significant differences in slope and absolute magnitude
are seen at higher and lower pulse energy fluxes.  This disagreement
can be understood by a combination of differences in antenna
sensitivity and pulse algorithm efficiency.  As discussed earlier, for
any fixed pulse energy a wide range of pulse amplitudes and widths are
observed.  A loss of pulse detection efficiency at low and high widths
may alter the true nature of the cumulative energy flux distribution.

Although earlier work by Bhat et al. \citep{bhat2008} only covered
widths from 1 to 10 $\mu$s, others \citep{popov2007} found a
dependence on pulse width, such that at narrow widths ($\sim$4
$\mu$s), $\alpha$=-1.7 whereas for wider pulses (65 $\mu$s), $\alpha$
was found to be =-3.2.  To examine this dependence our tight data set
was divided up into two subsets: a narrow ($W_{s}$ less than 5 $\mu$s)
and a wide data set ($W_{s}$ greater than 5 $\mu$s).  In both cases,
the slope of the cumulative energy flux distribution was estimated.
The narrow data set yields a slope values of -1.7 to -2.0, dependent
on the energy flux range fit, which is very consistent with Bhat et
al. While the wide data set yields a steep slope of $\alpha$=-2.4 to
-3, consistent with the work of Popov and Stappers.  

It is interesting to compare the total energy flux emitted by the Crab
pulsar as giant pulses with the overall pulse emission at this
frequency.  \ Using the background-free data set (tight cuts), we
summed the energy flux of each candidate GP, obtaining an average
pulse energy flux of 0.134 kJy-$\mu\mathrm{s}$ per rotation over the
span of the observation, which amounted to $\sim$ 168,000 rotations.
On the other hand, the mean pulse energy flux from the pulse profile containing all rotations
is 0.26 kJy--$\mu\mathrm{s}$.   This value
is in excellent agreement with the quoted value in the 
ATNF catalog\footnote{ATNF pulsar catalog web address:\\
http://www.atnf.csiro.au/research/pulsar/psrcat}
after extrapolating the mean flux density from 1400 MHz
to 1665 MHz using their spectral index of -3.1.
This suggests that a significant fraction, at least $\sim$50\%, of the pulsar
emission energy may be emitted as GPs. 
Another estimate for pulse energy flux obtained at a frequency of
1664 MHz, nearly identical to our observing frequency, suggests 
a mean energy flux of 0.125 kJy-$\mu\mathrm{s}$ for the main 
pulse and 0.025 kJy-$\mu\mathrm{s}$ for 
the interpulse at this frequency \citep{manchester1971}.
The total energy flux amounting to 0.15 kJy-$\mu\mathrm{s}$ is very close to the 
entire GP energy flux emission, further suggesting that a significantly
large portion of the energy flux may be emitted as GPs.   

\begin{figure}[h!t]
\centerline{\epsfig{file=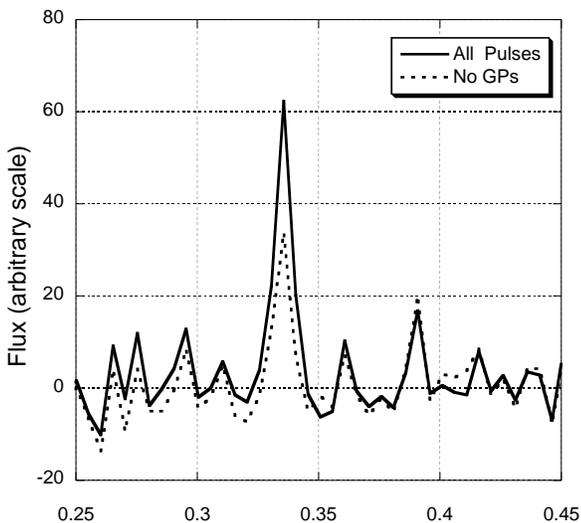, width=\linewidth}}
\caption{Crab pulse profiles obtained (solid curve) using the entire
sample discussed in this analysis.  The dashed curve shows the
profile without rotation cycles that include a GP candidate. 
}
\label{fig:profiles}
\end{figure}

As a further check we compared the pulse profile including all
rotations with a second profile, where rotations containing an
identified GP is removed from the profile.  Figure \ref{fig:profiles}
shows the two pulse profiles.  The second profile (dashed curve) is
obtained using a loose cut to identify GP candidates.  The figure
shows that GPs make up 54\% of the overall pulse energy flux at our
observing frequency.  Also, evident from the figure is how identical
the two profiles are in terms of the pulse width.  At the moment we do
not have sufficient statistics to resolve and quantify the width of
each profile and instead rely on a rough visual estimate.  This is
further indication that a similar significant reduction in energy flux
is obtained assuming that the GP and main pulse widths are similar.

A large fraction of the emission energy flux at this frequency is
already accounted for with GPs identified at our current sensitivity.
If the power law continues to lower energies, further improvement in
sensitivity will allow us to reach the intrinsic turnaround in the
distribution necessary to keep the total GP emission energy below the
total pulsed energy.  In this case, there would be only one
fundamental process responsible for the total pulsed emission in the
Crab pulsar.

\subsection { Pulse Asymmetry and Shape }

Skewness can be used as a measure of GP asymmetry, and is defined here to be:
\begin{equation}
s=\Sigma W_i(t_i- t_p)^3/(\Sigma W_i)\sigma^3,
\end{equation} 
where the weighted sum is over a narrow region of the pulse with
respect to its peak position $t_p$.  The weight, $W_i$, is the power
amplitude level in the corresponding time bin, $\sigma$ is the
standard deviation ($W_{std}$) and the pulse algorithm used is the
same as that used for obtaining the pulse standard deviation.  For a
Gaussian shaped pulse, skewness will have a value near zero.  Pulses
with long tails following the peak will have a positive skewness and
those with tails before the peak will have negative values.

\begin{figure}[h!t]
\centerline{\epsfig{file=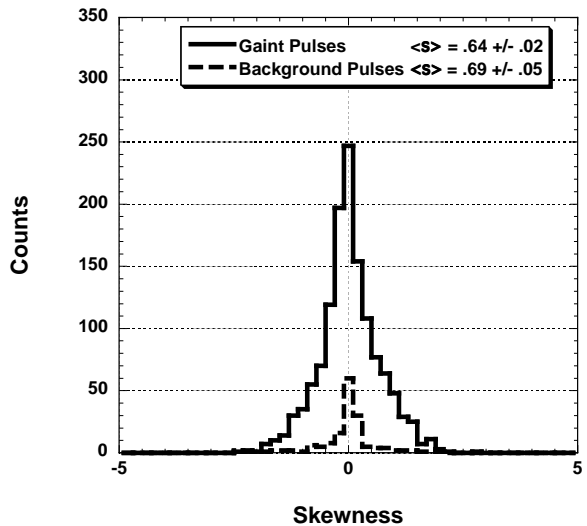, width=\linewidth}}
\caption{Skewness histograms for both the giant-pulses and
background pulses for the tight data set.}
\label{fig:skewness}
\end{figure}

Figure \ref{fig:skewness} shows the skewness distribution for both the giant 
pulses and background pulses, solid and dashed histogram respectively. 
Both distributions show average skewness that is very close to zero with 
very similar spread in the distribution.  
No statistically strong skewness dependence
was found as a function of width, amplitude or energy and no
differences were found between the interpulse and main giant pulses.

The next higher statistical moment, kurtosis, can also yield useful
information on pulse shape, and is defined as:
\begin{equation}
k=\Sigma W_i(t_i- t_p)^4/(\Sigma W_i)\sigma^4 - 3.
\end{equation}
Kurtosis is zero for a Gaussian-shaped pulse, negative for pulses that
are flatter than Gaussian, and positive for pulses more steeply peaked
than Gaussian.  To account for algorithmic dependences and background
noise, we have calculated skewness and kurtosis for both giant pulses
and out-of-phase background pulse-candidates.

\begin{figure}[h!t]
\centerline{\epsfig{file=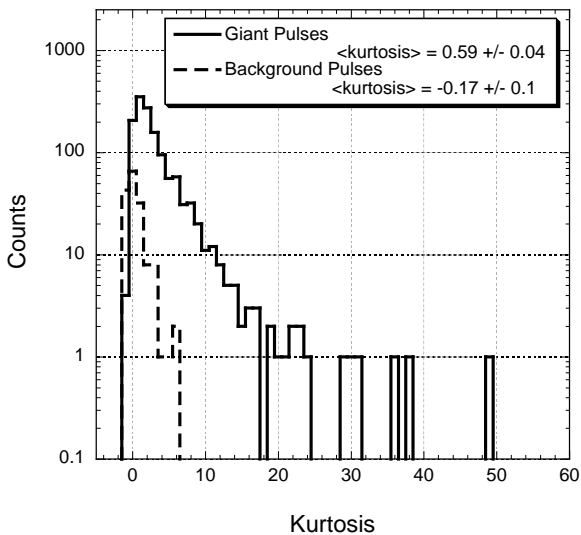, width=\linewidth}}
\caption{Histograms of pulse kurtosis for both giant pulses and
background pulses for the tight data set.}
\label{fig:kurtosis}
\end{figure}

\begin{figure}[h!t]
\centerline{\epsfig{file=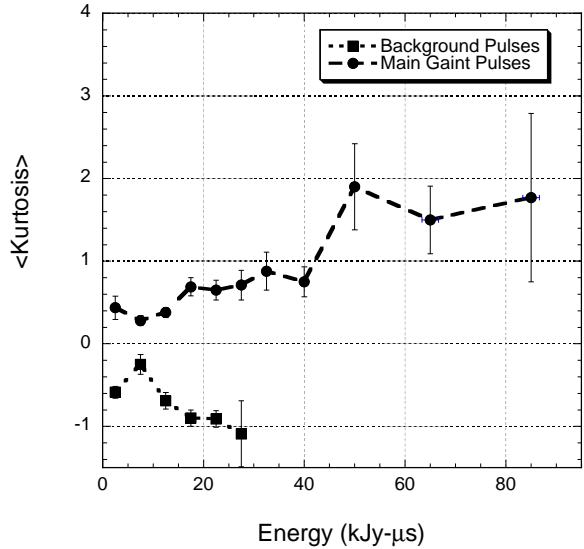, width=\linewidth}}
\caption{The dependence of kurtosis mean with the total pulse energy flux for
the giant pulses in the tight data set.}
\label{fig:energy_kurtosis}
\end{figure}

Our kurtosis distribution is shown in Figure \ref{fig:kurtosis}, for
both GPs and background events.  The background pulses are peaked near
zero, consistent with the expectation for gaussian shaped pulses. A
positive kurtosis tail extending to about a value of 5 is observed.
This most likely reflects the bias of the decimation portion of the
matched filter detection algorithm, where smoothing tends to increase
kurtosis.  The Crab's giant pulses are seen to be peaked at a kurtosis
value of ~0.6 with a very long tail extending to a value of 50,
implying they are more sharply peaked than pulses from the white-noise
background.  The reason for this can be seen by examining the
dependence of mean kurtosis on the total GP energy flux shown in
Figure \ref{fig:energy_kurtosis}.  This plot shows GP averaged
kurtosis increases linearly with the total energy flux.  Although the
background pulse candidates do not extend to high energies, it is
clear their behavior is quite different from the GPs upward trend.

\subsection{Pulse Time of Arrival}

\begin{figure}[h!t]
\centerline{\epsfig{file=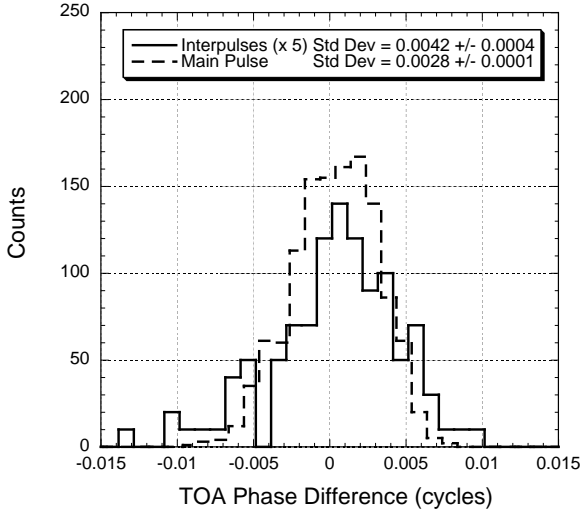, width=\linewidth}}
\caption{Histograms of the time of arrival for the Main and
Inter-pulse Giant pulses for the tight data set.}
\label{fig:toa_phase}
\end{figure}

The Crab pulsar has a well defined period of $\sim$33 ms and the
phase of the main and interpulse giant pulses are well regulated
within the pulsar period.  The phase of both the interpulse and main
pulses are seen in the projections of the scatterplot in Figure
\ref{fig:time_phase}.  The pulse phase residual, or Time of Arrival
(TOA) residual, is the TOA of the GP peak with respect the start of
the pulsar model cycle.  For the Crab, at L-band frequency, both the
main and interpulse GPs are found to have TOA residuals that fall
within 1$\%$ of a cycle ($\sim$ 330 $\mu$s).  Figure
\ref{fig:toa_phase} shows the histogram for the main and interpulse
GPs, where the mean TOA has been set to zero for both.  To remove
possible confusion in the interpulse data set due to background
pulse-candidates, the tight-cut data set has been used.  The main
pulses are found to have rms of 0.0028 $\pm$ 0.0001 cycles or
$1^\circ$ of phase corresponding to $\sim$ 90 $\mu$s, while the
interpulse GPs where found to be wider with an rms of 0.0042 $\pm$
0.0003 or 1.5$^\circ$ of phase corresponding to $\sim$ 140 $\mu$s.

Since it is expected that the intrinsic pulse phase jitter and
smearing will be independent of absolute phase location, the ~50$\%$
increase in TOA width is significant and may give an important
constraint in astrophysical pulsar models.  Two earlier studies, both
with smaller giant pulse samples, \citep{bhat2008,cordes2004} found
evidence for stronger pulses to have narrower phase windows. Where
strength was defined as the peak pulse amplitude.  Due to the strong
correlation of peak flux density and width, as shown in Figure 7a, it
is reasonable to believe that the narrow widths of the larger peak
pulses may indeed result in better phase resolution.  To examine this
possibility the tight data set was divided into two subsets, a large
and small peak flux density data set, those with peak amplitude larger
than 5 kJy and those smaller than 1 kJy. The widths of the TOA
residuals for the two n GP samples were then estimated to be $\sigma$=
0.0027 $\pm$ 0.0002 and $\sigma$= 0.0027 $\pm$ 0.0001 for the large
and small peak samples respectively. No significant difference is
found, as the sample TOA residuals are seen to be statistically
consistent to each other.

\begin{figure}[h!t]
\centerline{\epsfig{file=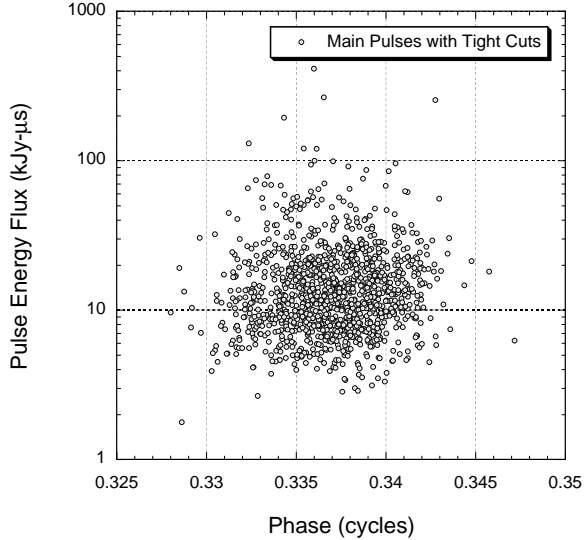, width=\linewidth}}
\caption{A scatterplot of the phase and total pulse energy flux.}
\label{fig:energy_phase}
\end{figure}

A plot of the main pulse energy flux versus phase is shown in Figures
\ref{fig:energy_phase}'s scatterplot.  No correlation between phase
and energy flux is evident.  To examine this more quantitatively, the
data were divided into three total-energy flux classes: high energy
flux (E $\geq$ 200 kJy-$\mu$s), medium energy flux (100 kJy-$\mu$s
$\leq$ E $\leq$ 200 kJy-$\mu$s), and low energy flux (E$\leq$100
kJy-$\mu$s).  The table below summarizes the TOA residual statistics
of each data set. No tendency is seen for larger energetic pulses to
originate in narrower phase windows.

\begin{deluxetable}{ccccc}
\footnotesize
\tablecaption{ Giant Pulse TOA residual statistics.\label{tbl-4}}
\tablewidth{0pt}
\tablehead{
  \colhead{E ( kJy-$\mu$s)} & \colhead{Sum} & \colhead{Mean} & \colhead{St.Dev.} & }

\startdata
E $\leq$ 100               & 564 & 0.3368 +/- 0.0001 & 0.0028 & \\
100 $\leq$ E $\leq$ 200  & 571 & 0.3373 +/- 0.0001 & 0.0028 & \\
E $\geq$ 200             & 106 & 0.3365 +/- 0.0003 & 0.0029 & \\

\enddata
\end{deluxetable}

\begin{figure}[h!t]
\centerline{\epsfig{file=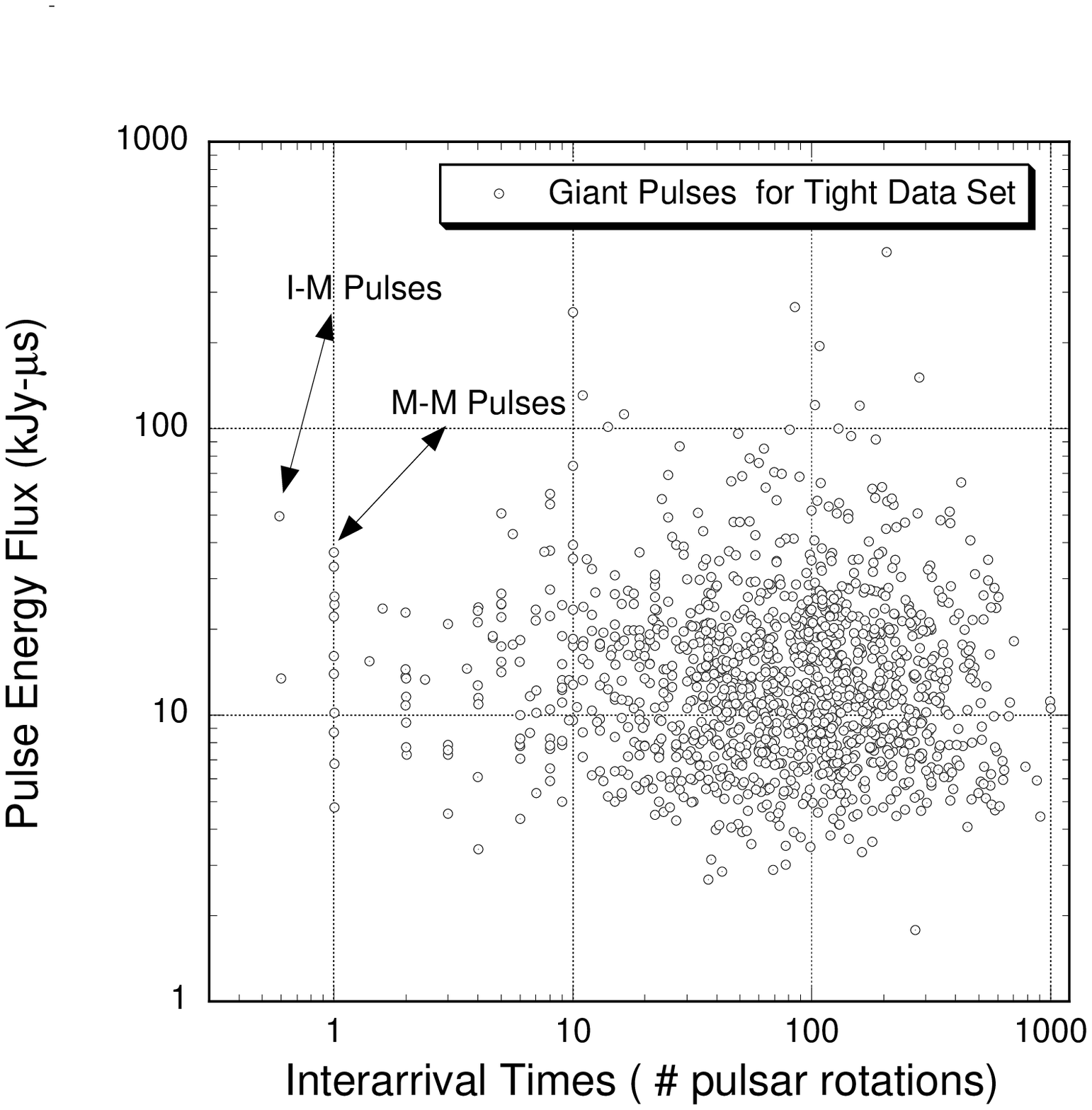, width=\linewidth}}
\caption{A scatterplot of the Giant-Pulse energy and and Interarrival
times for the Giant-Pulses in the tight data set.}
\label{fig:energy_iat}
\end{figure}

The joint statistics of GP total energy flux with the interarrival
time (IAT) of the next GP is shown in Figure \ref{fig:energy_iat}.
The IAT is simply the measured time between between one GP and the
next.  The interarrival time for both the Main pulse (MP) and
Interpulse (IP) giant pulses are plotted, where the IAT is converted
to the number of crab pulsar rotations ($\sim$ 33 ms).  The discrete
nature of the IAT for pulses below 10 rotations is simply a reflection
of the phase cut (0.30-0.35) for the MPs and (0.72-0.75) for the IPs.
For the sensitivity achieved with our observations, with tight cuts,
the average number of Crab rotations between giant pulses was found to
be 127 with the largest gap between giant pulses at $\sim$ 1000
rotations.  With the loosest cut used, the averaged number of Crab
rotations between GPs are 67.  No observable dependence between the
total GP energy flux and the arrival time of the next pulse was seen.

The smallest IAT of $\sim$ 0.4 cycle was observed twice, these cases
being a main pulse immediately preceded by an interpulse (I-M).  No
cases of interpulse immediately following the main pulse were found in
the tight data set (M-I), corresponding to an IAT of $\sim$ 0.6 cycle.
To examine the independence between main and interpulses, we examined
the consistency of the data with the assumed Poisson nature of the
giant pulse process.  A Poisson process describes events which occur
continuously and independently, and for this study would imply the
interarrival times are exponentially distributed with parameter
$\lambda$ (mean = 1/$\lambda$).

\begin{equation}
P(\tau)=\lambda e^{-\lambda \tau }\end{equation}

\begin{figure}[h!t]
\centerline{\epsfig{file=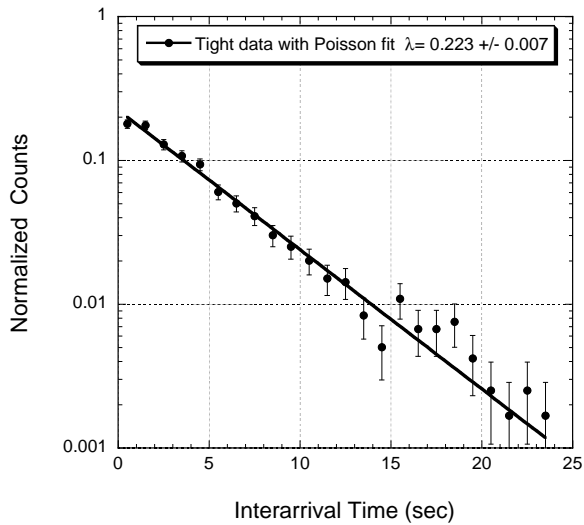, width=\linewidth}}
\caption{The interarrival time of main pulses (MPs) for the tight
 data set. The solid curve is an exponential fit to the data.}
\label{fig:iat}
\end{figure}

Figure \ref{fig:iat} shows the distribution of MPs interarrival times
along with the best fit to a Poisson model.  The data appears to be
consistent with a Poisson process, the best weighted fit yielding
$\lambda$=0.223 +/- 0.007 Hz with a $\chi^{2}$=19 for $\nu$=23.  This
$\lambda$ value agrees well the inverse of the mean interarrival
times, 0.234 +/- .007.  Within statistics, it appears that the arrival
times of Crab's giant pulses are Poisson distributed, implying they
are memoryless, and any given time interval is independent of what
occurs before or after.

In addition, both the interpulses and the combined GP data sets show
interarrival time distributions that are also consistent with a
Poisson process.  We conclude that interpulse GPs do not appear to be
correlated with main pulses to our level of examination.

\section{Summary and Conclusions}

We have examined the energy flux distribution, timing, and statistical
properties of giant radio pulses from the Crab pulsar at 1700 MHz
using DSN's 70~m antenna at Goldstone, achieving a time resolution of
125 ns.  Our pulse detection was based strictly on the underlying
Chi-square statistics of the data, where we attempted to keep the
noise rate and event confidence levels constant and independent of
pulse width.  Our consistent accounting of the probability
distributions and non-Gaussian effects, in order to keep background
rates independent of pulse width, is an improvement over previous
analyses.

The statistical analysis of giant pulse population was carried out
using three significance cuts: a loose cut (R-cut = 0.0200), where we
obtained 1879 pulses, and a tight cut (R-cut = 0.0006) yielding 1314
pulses and the loosest cut studied (R-cut = 0.6000) which yielded over
2500 main GPs.  With a large number of giant pulses we were able to
study various statistical properties of GPs.  We have confirmed the
power-law nature of the peak pulse flux density distribution and have
obtained a power-law slope of $-2.2$, consistent with \citep{bhat2008}
results.  We observed giant pulses with widths up to 100 $\mu$s, 10
times larger than the maximum found by Bhat et. al.  Further, the
distribution of pulse cumulative energy fluxes also follows a
power-law distribution with a slope of -2.57 , consistent with Bhat et
al. \citep{bhat2008}.  We also looked for correlations between the
time of arrival of individual giant pulses and concluded the time of
arrival of each event is consistent with Poisson statistics.  Our
pulse detection sensitivity, high-time resolution, and consistent
statistical treatment of the data has allowed us to examine a weaker
population of giant pulses, narrowing the gap with "normal" pulses
from the Crab pulsar.

Our comparison of pulsed energy flux emission from GPs with the normal
pulsed emission has forced us to claim that a significant portion, perhaps as
large as $\sim$90\% but at least 50\%, of the emission at this
frequency is in the form of GPs.  With further improvement in
sensitivity and observing time, it should be possible to either
confirm the turnaround seen in the energy distribution or push it only
slightly lower before reaching the intrinsic turnaround in the
distribution.  Such a study could be carried out by a phased array
instrument such as the VLA, where the narrower beam will make it
possible to lower the nebular contribution to the overall system
temperature, providing better sensitivity to further probe the weaker
population of GPs as well as the remaining normal pulses in the
system.

\section*{Acknowledgements}
We thank the staff of DSS-14 who helped in acquiring the data for this
work.  This work was carried out at the the Jet Propulsion Laboratory,
California Institute of Technology, under a Research and Technology
Development Grant.  Copyright 2010.  All Rights Reserved.  US
Government Support Acknowledged.

\bibliographystyle{apj}
\bibliography{crab.v2.4}

\end{document}